\documentclass[10pt,superscriptaddress,prl,aps,twocolumn,showpacs,amsmath,amssymb]{revtex4-2}
\usepackage{graphicx}
\usepackage{caption, subcaption}
\usepackage{hyperref}
\usepackage[capitalise]{cleveref}
\usepackage{dcolumn}
\usepackage{bm}
\newcommand{\Glasgow}[1]
{\affiliation{SUPA School of Physics and Astronomy, University of Glasgow, Glasgow G12 8QQ, 
United Kingdom}}

\newcommand{\Mainz}[1]
{\affiliation{Institut f\"ur Kernphysik, Johannes Gutenberg-Universit\"at Mainz, D-55099 
Mainz,Germany}}

\newcommand{\Bonn}[1]
{\affiliation{Helmholtz-Institut f\"ur Strahlen- und Kernphysik, Universit\"at Bonn, D-53115 
Bonn, Germany}}

\newcommand{\Dubna}[1]
{\affiliation{Joint Institute for Nuclear Research, 141980 Dubna, Russia}}

\newcommand{\Pavia}[1]
{\affiliation{INFN Sezione di Pavia, I-27100 Pavia, Italy}}

\newcommand{\GWU}[1]
{\affiliation{The George Washington University, Washington, DC 20052-0001, USA}}

\newcommand{\LPI}[1]
{\affiliation{Lebedev Physical Institute, 119991 Moscow, Russia}}

\newcommand{\Halifax}[1]
{\affiliation{Department of Astronomy and Physics, Saint Mary’s University, Halifax, Nova 
Scotia B3H 3C3, Canada}}

\newcommand{\Basel}[1]
{\affiliation{Departement f\"ur Physik, Universit\"at Basel, CH-4056 Basel, Switzerland}}

\newcommand{\Tomsk}[1]
{\affiliation{Laboratory of Mathematical Physics, Tomsk Polytechnic University, 634034 
Tomsk, Russia}}

\newcommand{\York}[1]
{\affiliation{Department of Physics, University of York, Heslington, York, Y010 5DD, UK}}

\newcommand{\INR}[1]
{\affiliation{Institute for Nuclear Research, 125047 Moscow, Russia}}

\newcommand{\Sackville}[1]
{\affiliation{Mount Allison University, Sackville, New Brunswick E4L 1E6, Canada}}

\newcommand{\Regina}[1]
{\affiliation{University of Regina, Regina, Saskatchewan S4S 0A2, Canada}}

\newcommand{\Zagreb}[1]
{\affiliation{Rudjer Boskovic Institute, HR-10000 Zagreb, Croatia}}

\newcommand{\Kent}[1]
{\affiliation{Kent State University, Kent, Ohio 44242-0001, USA}}

\newcommand{\Amherst}[1]
{\affiliation{University of Massachusetts, Amherst, Massachusetts 01003, USA}}
\newcommand{\Bochum}[1]
{\affiliation{Institut f\"ur Experimentalphysik, Ruhr-Universit\"at , D-44780 Bochum, 
Germany}}

\newcommand{\UCLA}[1]
{\affiliation{University of California Los Angeles, Los Angeles, California 90095-1547, 
USA}}

\newcommand{\aE}{\alpha_{E1}}

\newcommand{\bM}{\beta_{M1}}

\begin{document}

\title{Measurement of Compton scattering at MAMI for the extraction of the electric and magnetic polarizabilities of the proton}%

\author{E.~Mornacchi}\email{emornacc@uni-mainz.de}\Mainz\\ 
\author{P.P.~Martel} \Mainz\\ \Sackville \\
\author{S.~Abt}\Basel \\
\author{P.~Achenbach}\Mainz \\
\author{P.~Adlarson}\Mainz \\
\author{F.~Afzal}\Bonn \\
\author{Z.~Ahmed}\Regina \\
\author{J.R.M.~Annand}\Glasgow \\
\author{H.J.~Arends}\Mainz \\
\author{M.~Bashkanov}\York \\
\author{R.~Beck}\Bonn \\
\author{M.~Biroth}\Mainz \\
\author{N.~Borisov}\Dubna \\
\author{A.~Braghieri}\Pavia \\
\author{W.J.~Briscoe}\GWU \\
\author{F.~Cividini}\Mainz \\
\author{C.~Collicott}\Mainz \\
\author{S.~Costanza}\Pavia \\
\author{A.~Denig}\Mainz \\
\author{A.S.~Dolzhikov}\Dubna \\
\author{E.J.~Downie}\GWU \\
\author{P.~Drexler}\Mainz \\
\author{S.~Fegan}\York \\
\author{S.~Gardner}\Glasgow \\
\author{D.~Ghosal}\Basel \\
\author{D.I.~Glazier}\Glasgow \\
\author{I.~Gorodnov}\Dubna \\
\author{W.~Gradl}\Mainz \\
\author{M.~G\"unther}\Basel \\
\author{D.~Gurevich}\INR \\
\author{L. Heijkenskj{\"o}ld}\Mainz \\
\author{D.~Hornidge}\Sackville \\
\author{G.M.~Huber}\Regina \\
\author{A.~K{\"a}ser}\Basel \\
\author{V.L.~Kashevarov}\Mainz \\ \Dubna \\
\author{S.J.D.~Kay}\Regina \\
\author{M.~Korolija}\Zagreb \\
\author{B.~Krusche}\Basel \\
\author{A.~Lazarev}\Dubna \\
\author{K.~Livingston}\Glasgow \\
\author{S.~Lutterer}\Basel \\
\author{I.J.D.~MacGregor}\Glasgow \\
\author{D.M.~Manley}\Kent \\
\author{R.~Miskimen}\Amherst \\
\author{M.~Mocanu}\York \\
\author{C.~Mullen}\Glasgow \\
\author{A.~Neganov}\Dubna \\
\author{A.~Neiser}\Mainz \\
\author{M.~Ostrick}\Mainz \\
\author{D.~Paudyal}\Regina \\
\author{P.~Pedroni}\Pavia \\
\author{A.~Powell}\Glasgow \\
\author{T.~Rostomyan} \altaffiliation[Now at: ]{Paul Scherrer Institut (PSI), CH-5232 Villigen PSI, Switzerland} \Basel \\
\author{V.~Sokhoyan}\Mainz \\
\author{K.~Spieker}\Bonn \\
\author{O.~Steffen}\Mainz \\
\author{I.~Strakovsky}\GWU \\
\author{T.~Strub}\Basel \\
\author{M.~Thiel}\Mainz \\
\author{A.~Thomas}\Mainz \\
\author{Yu.A.~Usov}\Dubna \\
\author{S.~Wagner}\Mainz \\
\author{D.P.~Watts}\York \\
\author{D.~Werthm\"uller} \altaffiliation[Now at: ]{Paul Scherrer Institut (PSI), CH-5232 Villigen PSI, Switzerland} \York \\
\author{J.~Wettig}\Mainz \\
\author{M.~Wolfes}\Mainz \\
\author{N.~Zachariou}\York \\

\collaboration{A2 Collaboration at MAMI}

\date{\today}

\begin{abstract}
A precise measurement of the differential cross-sections $d\sigma/d\Omega$ and the linearly polarized photon beam asymmetry $\Sigma_3$ for Compton scattering on the proton below pion threshold has been performed with a tagged photon beam and almost $4\pi$ detector at the Mainz Microtron.  The incident photons were produced by the recently upgraded Glasgow-Mainz photon tagging facility and impinged on a cryogenic liquid hydrogen target, with the scattered photons detected in the Crystal Ball/TAPS set-up.  Using the highest statistics Compton scattering data ever measured on the proton along with two effective field theories (both covariant baryon and heavy-baryon) and one fixed-$t$ dispersion relation model, constraining the fits with the Baldin sum rule, we have obtained the proton electric and magnetic polarizabilities with unprecedented precision:
\begin{align*}
	&{}\alpha_{E1} = 10.99 \pm 0.16 \pm 0.47 \pm 0.17 \pm 0.34 \\
	&{}\beta_{M1} = 3.14 \pm 0.21 \pm 0.24 \pm 0.20 \pm 0.35
\end{align*}
in units of $10^{-4}$\,fm$^3$ where the errors are statistical, systematic, spin polarizability dependent and model dependent.\end{abstract}

\maketitle

\section{\label{sec:Intro} Introduction}

The study of hadron structure in terms of quantum chromodynamics and the
underlying quarks and gluons is a major focus of modern physics.  Due to the
nature of confinement and the complex internal dynamics involved, however, QCD
calculations of hadron properties have proved challenging.  The recent proton
radius ``puzzle''~\cite{Pohl:2013,Gao:2021} and the many measurements and theoretical
developments it has spurred, have emphasized that, while the proton is one of
the basic building blocks of matter and the most familiar of all hadrons, we
still do not fully understand its properties and structure.  Advances in effective field
theories~\cite{Lensky:2010,Griesshammer:2012,Lensky:2014,Lensky:2015}, dispersion relation
analyses~\cite{Drechsel:1999,Holstein:1999,Drechsel:2003,Pasquini:2007}, and lattice QCD~\cite{Detmold:2006}
have added impetus to obtain more accurate measurement of hadron structure observables, such as
polarizabilities and charge radii.

An object's polarizabilities characterize its internal response to applied external
electric ($\vec{E}$) and magnetic ($\vec{H}$) fields; they are fundamental properties such as mass and the charge and at the microscopic level, they can be accessed via Compton scattering.  
In the expansion of the effective Hamiltonian in incident photon energy $\omega_\gamma$, the electric ($\aE$) and magnetic ($\bM$) polarizabilities enter at $\mathcal{O}(\omega_{\gamma}^2)$~\cite{Babusci:1998}:
\begin{equation}
	H^{(2)}_{eff} = -4\pi \Bigl[\frac{1}{2} \alpha_{E1} \vec{E}^2 + \frac{1}{2} \beta_{M1} \vec{H}^2 \Bigr];
	\label{eq:H2}
\end{equation}
while the four spin polarizabilities ($\gamma_{EiMj}$) are included at $\mathcal{O}(\omega_{\gamma}^3)$:
\begin{align*}
	H^{(3)}_{eff} &= -4\pi \Bigl[ \frac{1}{2} \gamma_{E1E1} \vec{\sigma} \cdot (\vec{E} \times \dot{\vec{E}}) + \frac{1}{2} \gamma_{M1M1} \vec{\sigma} \cdot (\vec{H} \times \dot{\vec{H}}) \\
	&\quad- \gamma_{M1E2} E_{ij} \sigma_{i}H_{j} + \gamma_{E1M2} H_{ij} \sigma_{i}E_{j} \Bigr],
	\stepcounter{equation}\tag{\theequation}\label{eq:H3}
\end{align*}
where $\vec{\sigma}$ are the proton’s Pauli spin matrices, $\dot{\vec{E}} = \partial_{t} \vec{E}$ and $E_{ij} = \frac{1}{2}(\nabla_i E_j + \nabla_j E_i$) are partial derivatives with respect to time and space, respectively.
Considerable experimental effort has been expended over the
last half century to obtain the scalar polarizabilities of the
proton~\cite{Baranov:1974,Federspiel:1991,MacGibbon:1995,Olmos:2001}, and recent measurements have
resulted in extractions of the proton's heretofore unknown individual spin
polarizabilities~\cite{Martel:2015,Paudyal:2020}.  Work has also been carried out on the
neutron's scalar polarizabilities, but due to the absence of a free-neutron
target their unambiguous extraction has proved more
challenging~\cite{Hornidge:2001,Kossert:2002,Schumacher:2005,Myers:2014}.

Polarizabilities are of interest, not only in the study of hadron structure
where they can provide input to the QCD puzzle, but also in other fields including precision atomic physics and astrophysics.
They yield an appreciable correction to the proton charge radius via the Lamb shift and
hyperfine structure~\cite{Drell:1967,Bernabeu:1983,Faustov:2000,Carlson:2011} and
influence neutron star properties~\cite{Bernabeu:1974}.  Moreover, a precise
determination of the proton scalar polarizabilities is integral to the
extraction of the proton spin polarizabilities, as the latter appear at higher
order in the expansion of the Compton scattering
Hamiltonian~\cite{Low:1954,GellMann:1954,Klein:1955,Petrunkin:1961,Babusci:1998}.

Furthermore, based solely on improved chiral effective field theory analyses and attempts to curate a statistically consistent database~\cite{Lensky:2010,Griesshammer:2012,Lensky:2014,Lensky:2015}, the Particle Data Group have recently adjusted their values of the proton scalar
polarizabilities~\cite{PDG:2014} without using new experimental data. This clearly demonstrates the necessity to obtain a new Compton scattering dataset with high statistical accuracy and low systematic errors in order to constrain the extraction
of the scalar polarizabilities.

The measurement reported here represents the highest statistics Compton
scattering data ever obtained on the proton --- roughly one million Compton
events below pion photoproduction threshold --- resulting in an improvement of a factor of
approximately five over the world's previous best measurement~\cite{Olmos:2001}.
It builds on our recently reported pilot experiment where the photon beam asymmetry below pion
threshold was obtained for the very first time~\cite{Sokhoyan:2017}. 

\section{\label{sec:Exp} Experimental Setup and Data Handling}

The experimental data were obtained in two beamtimes in 2018 using tagged photons at the MAMI electron microtron facility~\cite{Kaiser:2008,Jankowiak:2006}. 
The electron beam, with an energy of $883$~MeV, impinged on a $10~\mu$m thin diamond radiator, producing a linearly polarized photon beam via coherent Bremsstrahlung~\cite{Lohmann:1994}, with a degree of polarization up to $78\%$. 
The recoiling electrons from the Bremsstrahlung process were momentum analyzed using the Glasgow photon tagging spectrometer~\cite{McGeorge:2007}. 
Only photons with energies in the range $\omega_{\gamma} = 85 - 140$~MeV were considered in this analysis. 
The resulting photon beam passed a $3$-mm-diameter lead collimator and was incident on a $10$-cm-long liquid hydrogen target.
The final state particles were detected using the same Crystal Ball (CB)/TAPS detector system as in the pilot experiment~\cite{Sokhoyan:2017} with a nearly complete solid angle coverage. Additional information on the apparatus used for these measurements can be found in Refs.~\cite{Unverzagt:2008,Prakhov:2008}.%
\begin{figure}[]%
    \begin{center}%
        \includegraphics[width=0.45\textwidth,angle=0]{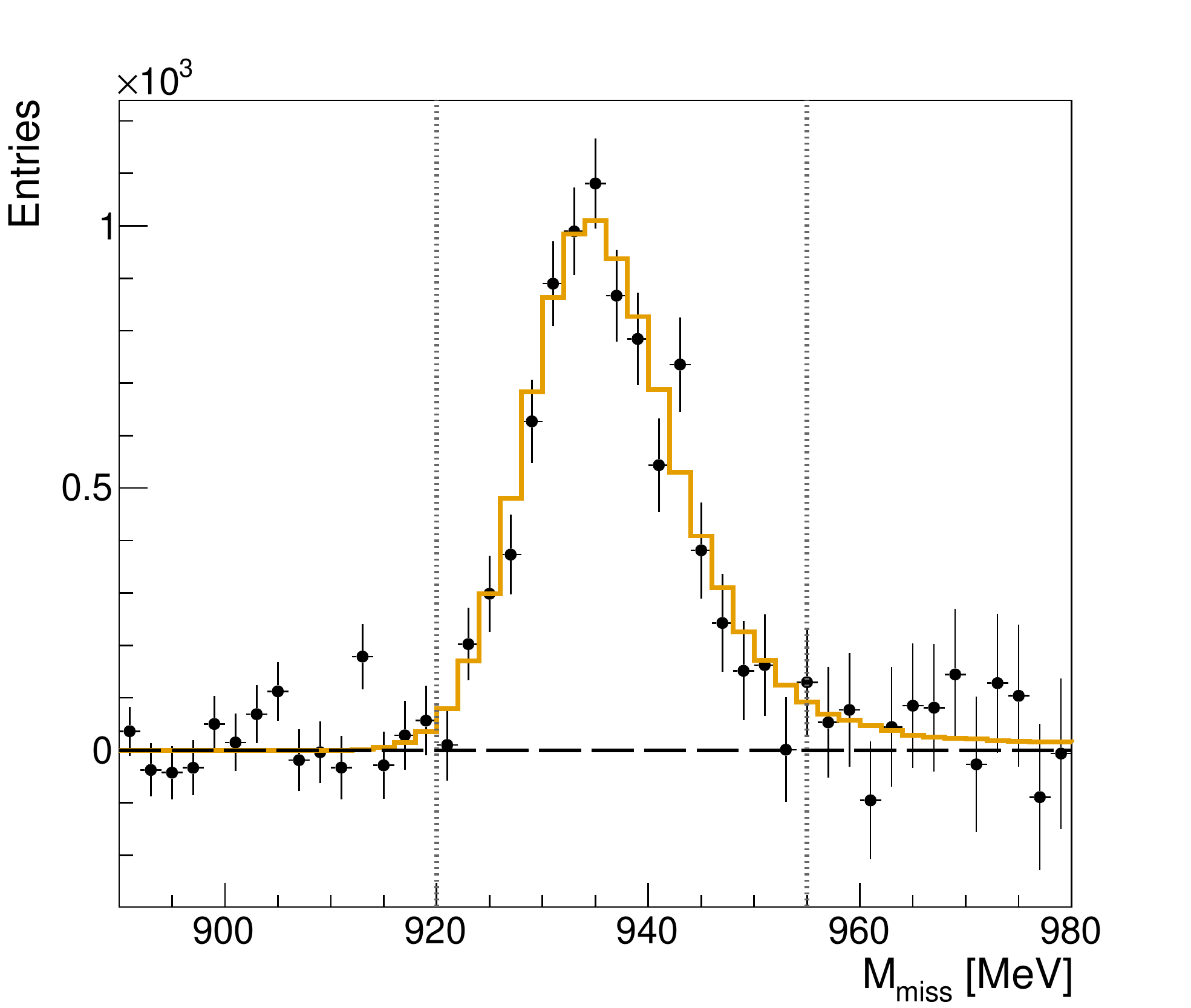}%
    \end{center}%
    \caption{Example of missing mass distribution for $\theta_{\gamma'} = 120-130^{\circ}$ and $\omega_{\gamma} = 118-130~\text{MeV}$. The measured missing mass distribution is shown together with the simulated Monte Carlo one, in black and orange, respectively. The gray dotted lines show the selection applied in the analysis.}
    \label{fig:missingmass}
\end{figure}
\begin{figure*}%
    \begin{center}%
    \subfloat[\label{fig:Results:a}Unpolarized differential cross-section.]{\includegraphics[width = 0.33\linewidth]{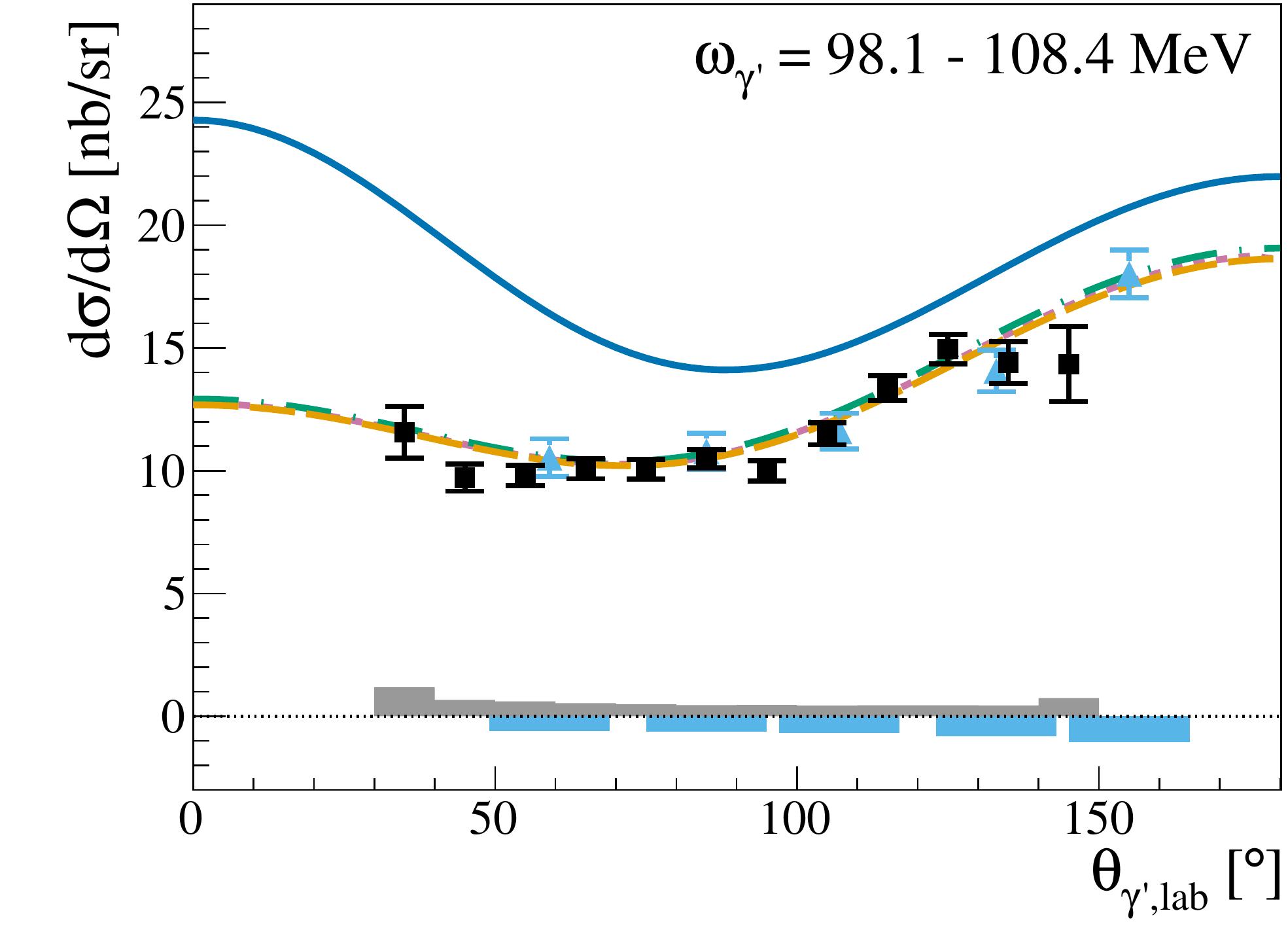} \includegraphics[width = 0.33\linewidth]{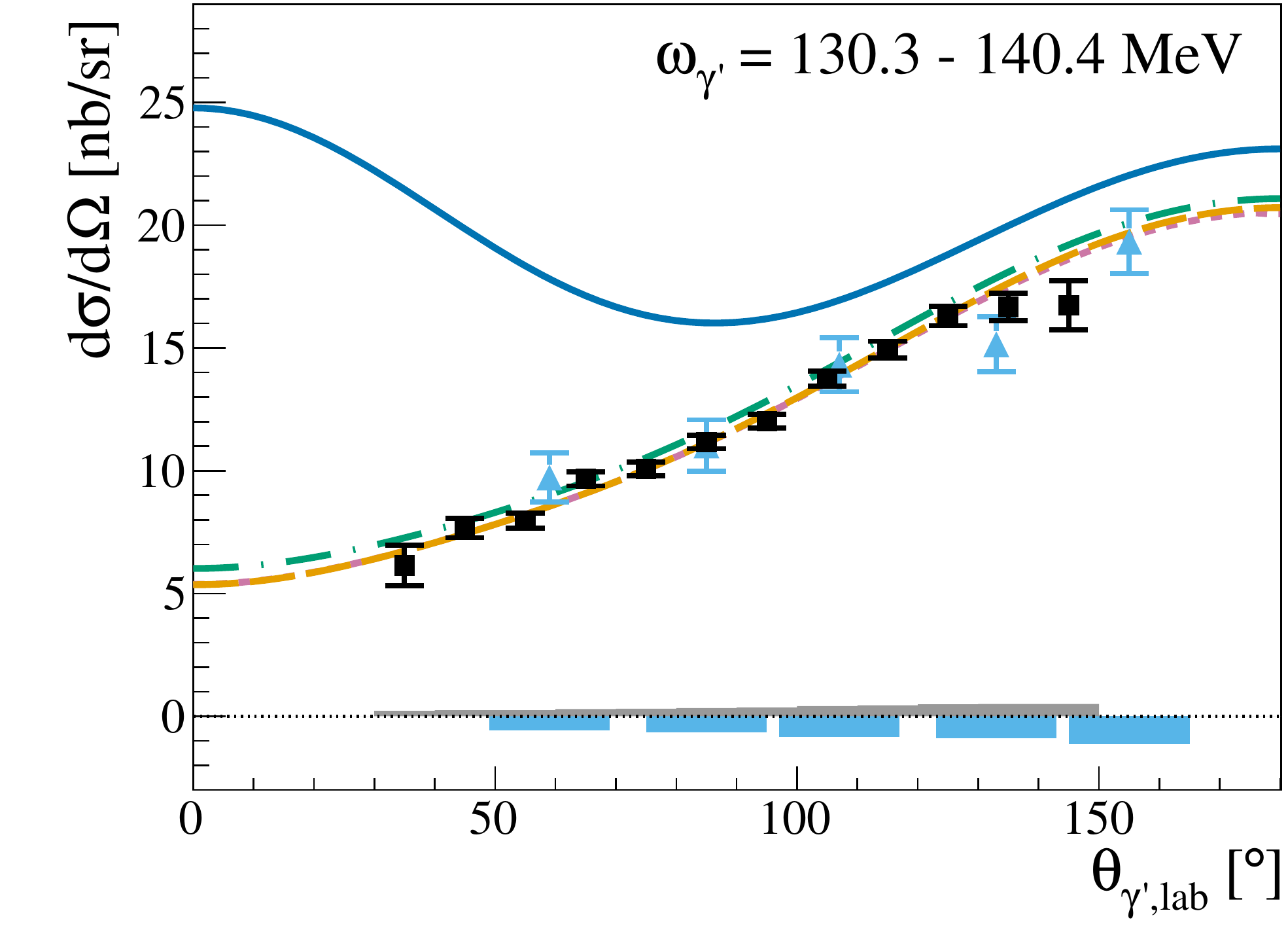}}%
	\subfloat[\label{fig:Results:b}Beam asymmetry.]{\includegraphics[width = 0.33\linewidth]{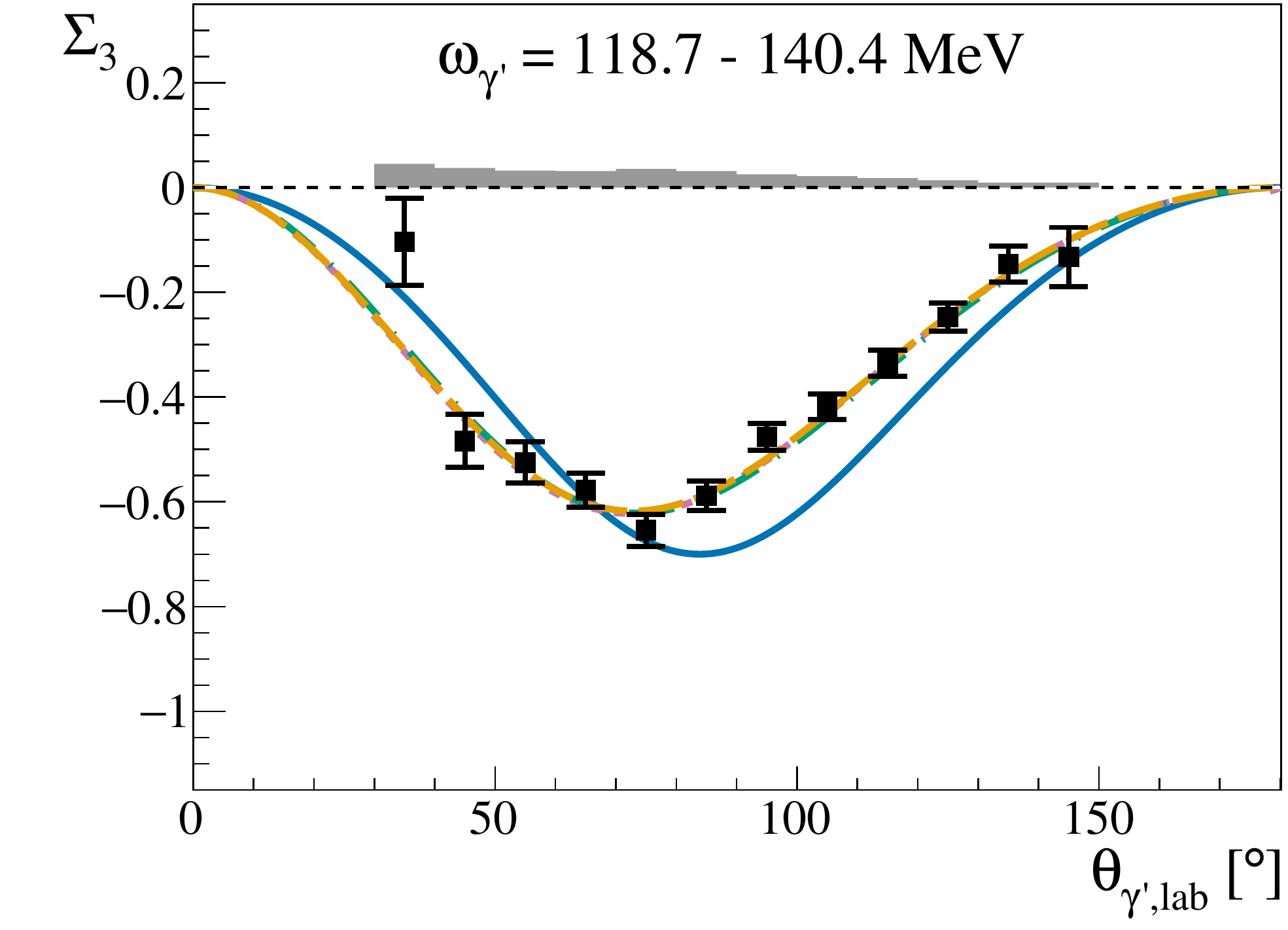}}%
 	\end{center}%
	\caption{Examples of our new data on the proton Compton scattering. Panel (a) shows 24 of the 60 total points of unpolarized differential cross-section. The azure triangles show previous measurement from the TAPS collaboration~\cite{Olmos:2001} for the closest energy bin ($\omega_{\gamma}$ = 98.9 and 134.7 MeV from left to right, respectively). Panel (b) shows 12 of the 36 total points of beam asymmetry $\Sigma_3$. In all the plots, the errors are statistical only. The systematic uncertainties are depicted as gray and azure bars for the new points and the TAPS results, respectively. The solid blue curves represent the Born contribution only. The long-dashed orange, short-dashed purple and dotted-dashed green curves represent our fit results reported in \cref{tab:results}, obtained within DR~\cite{Drechsel:1999,Holstein:1999,Pasquini:2007}, B$\chi$PT~\cite{Lensky:2010} and HB$\chi$PT~\cite{McGovern:2012} frameworks, respectively. All the measured points are reported and plotted in the Supplemental Material~\cite{SuppMat}}%
	\label{fig:Results}%
\end{figure*}%

The experiment was performed with two different orthogonal orientations of the photon polarization vector (formed by the momentum of the incoming photon and its electric field vector) and to minimize the systematic uncertainty the polarization vector was flipped approximately every two hours. 
The degree of linear polarization, that depends on the photon energy and crystal orientation, was directly extracted from experimental data following the procedure described in Ref.~\cite{Livingston:2008}.

The Compton scattering events, $\vec{\gamma} p \to \gamma p$, were selected by requiring exactly one photon in the CB (for the incident photon energies of interest the recoil protons are typically not energetic enough to reach the detector apparatus).
The photon is identified as a cluster in the calorimeter without any associated hit in the Particle Identification Detector nor in any of the two Multiwire Proportional Chambers. 
Due to a significant beam-related electromagnetic background in the forward region, only events with outgoing photon scattering angle $\theta_{\gamma'} > 30^{\circ}$ were considered.
Moreover, due to the relatively high photon flux, a time coincidence within $3$~ns was required between the scattered photon and the hits in the tagger focal plane. 
To remove the random coincidences in the selected time window, a side-band subtraction was also performed by selecting a background sample on each side of the prompt peak.
Furthermore, the background generated from the the non-hydrogenic components of the target (Kapton cell, insulating material, etc.) was sampled during dedicated empty-target runs, and subtracted from the full-target sample.

\cref{fig:missingmass} shows a sample of the missing mass distribution calculated as
 \begin{equation}
     	M_{miss} = \sqrt{(\omega_{\gamma} + m_p  - \omega_{\gamma'})^2 - (\vec{k} - \vec{k}')^2 },
 	\label{eq:missingmass}
 \end{equation}
where $k = (\omega_{\gamma}, \vec{k})$ and $k' = (\omega_{\gamma'}, \vec{k'})$ are the incoming and scattered photon four-momenta, respectively, and $m_p$ is the target proton mass at rest. 
For a Compton scattering event from a proton in the target cell we expect $M_{miss}$ to be in agreement with the proton mass.
The distribution is plotted together with the one obtained from a Monte Carlo simulation of the full experimental apparatus, based on the Geant4 package~\cite{GEANT4:2002}.
The good agreement between the data and the simulated distribution indicates a low background contamination in the final sample, and to remove any remaining background a cut on missing mass is applied (\cref{fig:missingmass}), which was optimized using the Monte Carlo simulation.

From the final dataset, the unpolarized differential cross-section was extracted for five approximately $10$~MeV-wide photon energy bins spanning the range $\omega_{\gamma} = 86$ to $140$~MeV and twelve $10^{\circ}$-wide polar angular bins from $\theta_{\gamma'} = 30^{\circ}$ to $150^{\circ}$, for a total of 60 points. 
A sample of the obtained results is plotted in \cref{fig:Results:a} as the statistical-error-weighted average between the two beamtimes.
The error bars are statistical only with the systematic uncertainty given by gray bars. They include both correlated and point-to-point uncorrelated errors. 
The former comes from three independent sources that give in total a systematic uncertainty of $3\%$: target density ($1\%$), photon flux normalization ($2\%$), and analysis cuts and Monte Carlo simulation ($2\%$). 
The latter comes from the remaining background contamination in the final sample and it was estimated from the small yield evident to the right of the proton peak before the missing mass cut (\cref{fig:missingmass}).
This background yield was found to be energy- and polar-angle-dependent and ranges from $10\%$ at low beam energy and scattering angle to $0.2\%$ at high energy and central angular bins.
The consistency of the two beamtimes was checked looking at the distribution of the normalized residuals of the two cross-sections calculated using the two different periods. 
The results were found to be in perfect agreement for all energy bins, but the lowest one ($\omega_{\gamma} = 86.3 - 98.2~\text{MeV}$) in which one of the two beamtimes was about $5\%$ higher than the other one.
To account for this, a $3\%$ systematic uncertainty was included in the uncorrelated systematic uncertainties of this lowest energy bin.
The numerical values are tabulated and plotted in the Supplemental Material~\cite{SuppMat}.
Our results in \cref{fig:Results:a} are compared with the previously published data from the TAPS collaboration~\cite{Olmos:2001}.  
The Born term, describing the proton as a point-like particle without internal structure aside from the anomalous magnetic moment, is shown along with the fit results reported in \cref{tab:results}, obtained within different theoretical frameworks: Dispersion relation (DR)~\cite{Drechsel:1999,Holstein:1999,Pasquini:2007}, Baryon Chiral Perturbation theory (B$\chi$PT)~\cite{Lensky:2010}, and Heavy Baryon Chiral Perturbation theory (HB$\chi$PT)~\cite{McGovern:2012}.
The main difference between the two $\chi$PT variants is whether the nucleon is treated relativistically (B$\chi$PT) or an amplitude expansion in powers of $1/M_{N}$ is performed (HB$\chi$PT) (see Ref.~\cite{Lensky:2012} for a comprehensive comparison).
The three theories fit our data equally well, as confirmed also from the $\chi^2$ values in \cref{tab:results}.
The experimental data show divergence from a calculation based on only the Born term, showing the sensitivity of the data to the proton internal structure, which at this energy is mainly, but not exclusively, described by the scalar polarizabilities.
The improvement in the statistical quality of the new data compared to the previous TAPS measurement is clear from \cref{fig:Results:a}. The new measurement also provides a wider angular coverage and smaller systematic errors.

The combination of the linearly polarized photon beam with the unpolarized LH$_2$ target results in azimuthal dependence to the Compton scattering cross-section, allowing for the determination of the single polarization observable $\Sigma_3$ defined as~\cite{Babusci:1998}
 \begin{equation}
     \Sigma_3 = \dfrac{d\sigma_{\parallel} - d\sigma_{\perp}}{d\sigma_{\parallel} + d\sigma_{\perp}},
     \label{eq:sigma3}
 \end{equation}
where $d\sigma_{\parallel(\perp)}$ is the polarized cross-section obtained with one of the two orthogonal orientations of the photon polarization vector, usually named ``parallel'' and ``perpendicular''.

The beam asymmetry was extracted in the same energy and angular range as the unpolarized cross-section, using a procedure similar to the one in Ref.~\cite{Sokhoyan:2017}. Due to the larger statistical uncertainties, to achieve adequate statistics the current data were binned in three photon energy regions, for a total of 36 new points.
A sample of our results is shown in \cref{fig:Results:b}.
The error bars shown are statistical only. The systematic uncertainties are depicted as gray bars. 
The main contribution to the systematic uncertainty comes from the procedure to extract the linear polarization degree from the data, and an upper limit to this is estimated to be $5\%$, uniformly distributed. 
An additional uncorrelated point-to-point contribution coming from the background contamination was also estimated, using a method similar to that employed for the unpolarized cross-section.
The numerical values are tabulated and plotted in the Supplemental Material~\cite{SuppMat}.
Our asymmetry results in \cref{fig:Results:b} are plotted together with the fit results obtained using the same theoretical frameworks as for the differential cross-sections shown in \cref{fig:Results:a}.
Also in this case, the three models can fit our results equally well.
However, in all three photon energy bins the fit results are distinct from the leading Born term, indicating that the asymmetry is sensitive to the proton structure constants. 

A more comprehensive description of this work can be found in Ref.~\cite{Mornacchi:2021}.

\section{\label{sec:Results} Results and Discussion}

\begin{table*}[]
\caption{Scalar polarizabilities extracted by fitting the new unpolarized cross-section and beam asymmetry data using HDPV DR~\cite{Drechsel:1999,Holstein:1999,Pasquini:2007}, B$\chi$PT~\cite{Lensky:2010}, and HB$\chi$PT~\cite{McGovern:2012} code. The errors are statistical, systematic, and from the spin polarizabilities, respectively. The spin polarizabilities were fixed to the last experimental values available~\cite{Paudyal:2020}. $s_{\sigma}$ and $s_{\Sigma}$ are the normalization factors for the unpolarized cross-section and the beam asymmetry, respectively. The scalar polarizability values are given in units of $10^{-4}~\text{fm}^3$.} \label{tab:results}
    \resizebox{\linewidth}{!}{%
    \begin{tabular}{|l|c|c|c|}
        \hline
                        & HDPV					    			& B$\chi$PT								& HB$\chi$PT \\ \hline
        $\aE$           & $11.23 \pm 0.16 \pm 0.46 \pm 0.02$ & $10.65 \pm 0.16 \pm 0.47 \pm 0.04$ & $11.10 \pm 0.16 \pm 0.47 \pm 0.17$ \\ 
        $\bM$           & $2.79  \pm 0.20 \pm 0.23 \pm 0.11$ & $3.28  \pm 0.21 \pm 0.24 \pm 0.09$ & $3.36  \pm 0.21 \pm 0.24 \pm 0.20$  \\ \hline
        $s_{\sigma}$    & $1.011 \pm 0.015$					& $1.013 \pm 0.015$         			& $1.043 \pm 0.016$           \\
        $s_{\Sigma}$    & $0.994 \pm 0.015$        	 		& $0.996 \pm 0.015$         			& $1.001 \pm 0.015$           \\ \hline
        $\chi^2$/DOF    & $82.10/93 = 0.89$         			& $82.96/93 = 0.89$         			& $83.16/93= 0.89$	\\ \hline
			\end{tabular}%
		}
\end{table*}
A fit to extract the proton scalar polarizabilities from all of the new data was performed using  three different models: fixed-t DRs~\cite{Drechsel:1999,Holstein:1999,Pasquini:2007}, B$\chi$PT~\cite{Lensky:2010}, and HB$\chi$PT~\cite{McGovern:2012}.
The unpolarized cross-section and the beam asymmetry were given as input to the fitter as two independent datasets.
The uncorrelated point-to-point systematic errors were added in quadrature to the statistical ones. 
The correlated systematic uncertainties were included in the fit as common normalization factors, one for each dataset, and treated as additional fit parameters. 
Their deviations from the expected value of 1 were also accounted for in the $\chi^2$ function to be minimized~\cite{DAgostini:1993}. 
The minimization was performed by using 
the MINUIT minimization routine~\cite{James:1975}.
In order to emphasize the sensitivity of the new data to $\aE$ and $\bM$ as much as possible, the spin polarizabilities were kept fixed to the most recent experimental values~\cite{Paudyal:2020}.
Moreover, to minimize the statistical uncertainty, the well-known Baldin Sum rule was included as an additional data point to be fitted at $\aE + \bM = 13.8 \pm 0.4$ (in the usual units)~\cite{Olmos:2001}.
\begin{figure}[]
    \begin{center}
        \includegraphics[width=0.45\textwidth,angle=0]{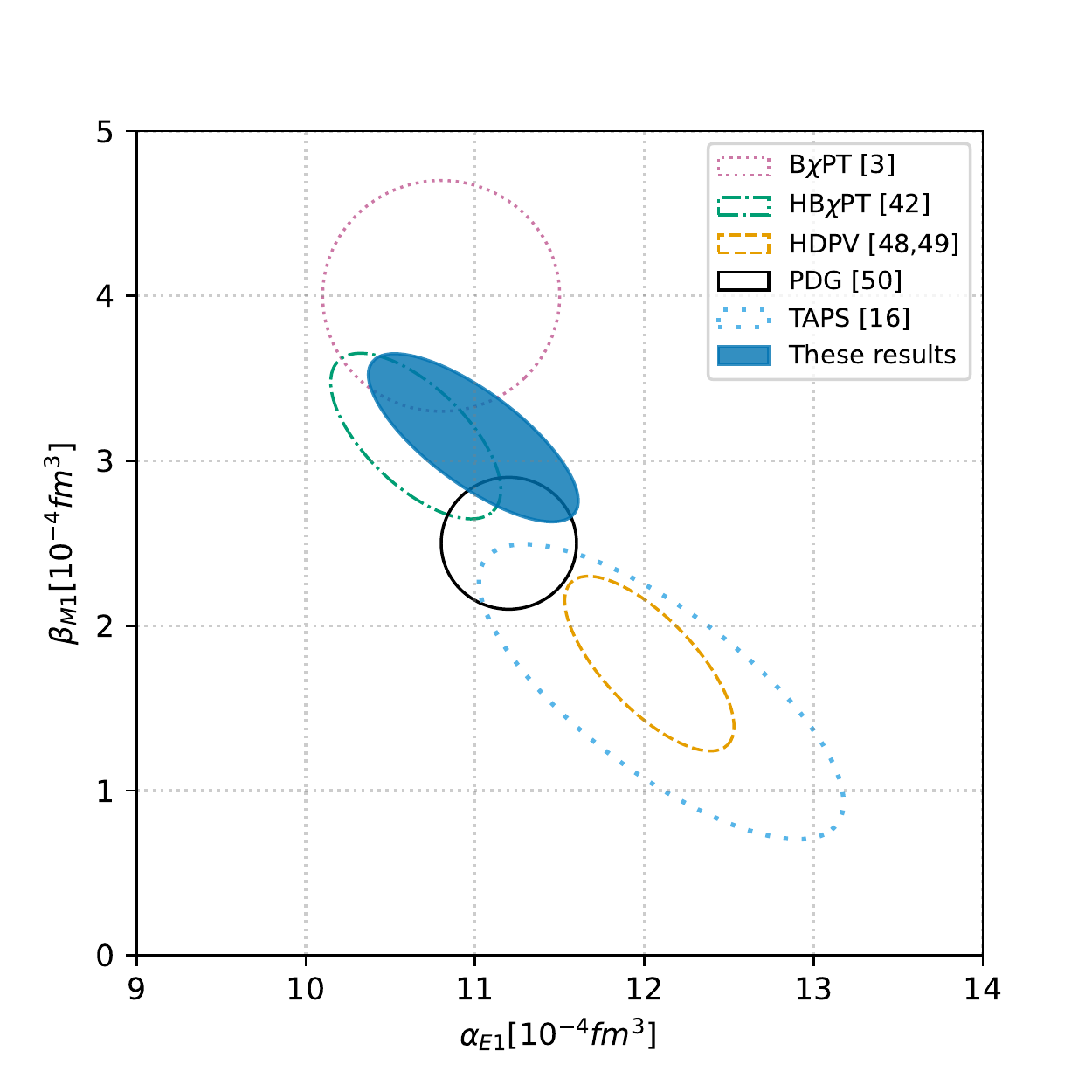}
    \end{center}
    \caption{Results of $\aE$ vs $\bM$ for the proton, obtained from different experiments and theories. The extraction from our data is depicted as blue full ellipse. The loosely dotted azure ellipse shows the result from the TAPS collaboration~\cite{Olmos:2001}. The dotted purple circle is the B$\chi$PT prediction~\cite{Lensky:2010}, the green dashed-dotted curve is the extraction within HB$\chi$PT~\cite{McGovern:2012}, and the orange dashed curve is the bootstrap-based fit using DR~\cite{Pasquini:2019, Pedroni:2019}. The black circle shows the values quoted by the PDG~\cite{Zyla:2020}. The Baldin sum rule constraint was used in the present extraction, as well as in those from TAPS, HB$\chi$PT, and HDPV.  All contours correspond to 1$\sigma$ level.}
    \label{fig:alphabeta2D}
\end{figure}

The fit results are summarized in \cref{tab:results}. 
The errors quoted in the central values of $\aE$ and $\bM$ are statistical, systematic and from the spin polarizabilities, respectively.
The first one was obtained by performing the fit without the normalization factors. 
The second error is given by how much the errors on the parameters changed by the inclusion of the systematic errors. 
The third error is given by the variation in the best value of $\aE$ and $\bM$ when the spin polarizabilities are not fixed, but rather free to vary within their experimental errors.
The small systematic error for the latter term indicates the new dataset has only a limited dependency on the spin polarizabilities, and thus making it well suited for a precise study of the two scalar terms.

The extractions of the scalar polarizabilities reported in \cref{tab:results} --- in particular of $\bM$ --- exhibits a moderate model dependence. 
To provide a best estimate of the central values for the two parameters, the results from the three theories were combined using weighted average, taking the quadratic sum of the statistical and systematic uncertainties as weights. For each error the largest contributions among the different theories was assigned.
Additionally, the largest of the differences between each theory and the average was used to estimate an additional error due to the model dependence for both $\aE$ and $\bM$. 
The best values for the extraction of the scalar polarizabilities from the new data using the Baldin sum rule constraint are
\begin{align*}
	&{}\aE = 10.99 \pm 0.16 \pm 0.47 \pm 0.17 \pm 0.34 \\
	&{}\bM = 3.14 \pm 0.21 \pm 0.24 \pm 0.20 \pm 0.35 \stepcounter{equation}\tag{\theequation}\label{eq:finalresults}
\end{align*}
where the errors are statistical, systematic, spin polarizability dependent and model dependent. 
A correlation coefficient between the two scalar polarizabilities of $\rho_{\aE-\bM} = -0.75$ was also reported by the fitter.
The effect of the constraint was checked by repeating the fits without the additional point at $\aE + \bM = 13.8 \pm 0.4$.
The obtained values for $\aE$ and $\bM$ are in agreement with the ones of \cref{eq:finalresults} within $1.5\sigma$ and $0.5\sigma$, respectively, indicating the limited effect of the constraint on the final results.

\Cref{fig:alphabeta2D} shows the scalar polarizability extraction from this work as the blue full ellipse. Also shown are various previously published global extractions and predictions of these two parameters.
The azure dotted circle shows in particular the results from the TAPS collaboration~\cite{Olmos:2001}, the highest statistics dataset published previously.
The improvement in the uncertainty of the scalar polarizabilities extracted from the new data is clearly visible.

\section{\label{sec:final} Summary}

In summary, a new precision measurement of the proton Compton scattering unpolarized cross-section and
beam asymmetry is presented. 
A fit to the new data using different theoretical models resulted in an extraction of the scalar polarizabilities $\aE$ and $\bM$ from one consistent dataset with an unprecedented precision. The new results will be important for resolving the current ambiguities in the extraction of these fundamental quantities.
Moreover, these new experimental data can be used in combination with the already published ones on single and double polarization observables from the A2 collaboration~\cite{Martel:2015,Sokhoyan:2017,Paudyal:2020}, to obtain the first combined extraction of all the six proton polarizabilities from experimental data measured at a single facility, achieving an important new milestone in the MAMI program.

\begin{acknowledgments}
The authors wish to acknowledge the outstanding support of the accelerator group and operators of MAMI. We also wish to acknowledge and thank J.~McGovern, V.~Pascalutsa, and B.~Pasquini for providing us with their theory codes, together with H.~Grie{\ss}hammer and M.~Vanderhaegen for the theoretical contributions and support. 
This project has received funding from the European Union’s Horizon 2020 research and
innovation program under grant agreement No 824093. 
This work has been supported by the U.K. STFC (ST/L005719/1, ST/P004458/1, ST/T002077/1,ST/P004385/2, ST/V002570/1, ST/P004008/1 and ST/L00478X/2) grants, the Deutsche Forschungsgemeinschaft (SFB443, SFB/TR16, and SFB1044), DFG-RFBR (Grant No. 09-02-91330), Schweizerischer Nationalfonds (Contracts No. 200020-175807, No. 200020-156983, No. 132799, No. 121781, No. 117601), the U.S. Department of Energy (Offices of Science and Nuclear Physics, Awards No. DE-SC0014323, DEFG02-99-ER41110, No. DE-FG02-88ER40415, No. DEFG02-01-ER41194) and National Science Foundation (Grants NSF OISE-1358175; PHY-1039130, PHY-1714833, PHY-2012940 No. IIA-1358175), INFN (Italy), and NSERC of Canada (Grant No. FRN-SAPPJ2015-00023).
\end{acknowledgments}

\end{document}